\documentclass[10pt,twocolumn,letterpaper]{article}

\usepackage[pagenumbers]{cvpr} 
\usepackage{multirow}

\usepackage{verbatim}
\usepackage{amsmath,amssymb,amsthm}
\usepackage{graphicx}
\usepackage{grffile}
\usepackage{epstopdf}
\usepackage{multirow}
 \usepackage{verbatim}
\usepackage{verbatim}
\usepackage{makecell}
\usepackage{pifont}
\usepackage{amssymb}
\usepackage{float}
\usepackage{graphicx}
\usepackage{multirow}

\usepackage{url}
\usepackage{textcomp}
\usepackage{epstopdf}
\usepackage{algorithm}
\usepackage{algpseudocode}
\usepackage{booktabs}
\usepackage{cases}
\usepackage{caption}
\usepackage{float}
\usepackage{threeparttable}
\usepackage{makecell}
\usepackage{placeins}

\usepackage[dvipsnames]{xcolor}

\definecolor{cvprblue}{rgb}{0.21,0.49,0.74}
\usepackage[pagebackref,breaklinks,colorlinks,citecolor=cvprblue]{hyperref}

\def\paperID{} 
\def\confName{}
\def\confYear{}

\title{DSFNet: Dual-GCN and Location-fused Self-attention with Weighted Fast Normalized Fusion for Polyps Segmentation}

\newcommand{\hs}[0]{\hspace*{0.7cm}}

\author{\noindent\parbox{0.97\textwidth}{
\centering
Juntong Fan\textsuperscript{1}\hs
Debesh Jha\textsuperscript{2}\hs
Tieyong Zeng\textsuperscript{1}\hs
Dayang Wang\textsuperscript{3}\\ \vspace{0.5em}
$^1$Department of Mathematics, The Chinese University of Hong Kong, Hong Kong. \\ 
$^2$Department of Radiology, Northwestern University, Evanston, IL, US \\ 
$^3$Department of Electronic and Computer Engineering, University of Massachusetts, Lowell, Lowell, MA, USA.}
}

\begin{document}
\maketitle
\begin{abstract}
Polyps segmentation poses a significant challenge in medical imaging due to the flat surface of polyps and their texture similarity to surrounding tissues. This similarity gives rise to difficulties in establishing a clear boundary between polyps and the surrounding mucosa, leading to complications such as local overexposure and the presence of bright spot reflections in imaging. To counter this problem, we propose a new dual graph convolution network (Dual-GCN) and location self-attention mechanisms with weighted fast normalization fusion model, named DSFNet. First, we introduce a feature enhancement block module based on Dual-GCN module to enhance local spatial and structural information extraction with fine granularity. Second, we introduce a location fused self-attention module to enhance the model's awareness and capacity to capture global information. Finally, the weighted fast normalized fusion method with trainable weights is introduced to efficiently integrate the feature maps from encoder, bottleneck, and decoder, thus promoting information transmission and facilitating the semantic consistency. Experimental results show that the proposed model surpasses other state-of-the-art models in gold standard indicators, such as Dice, MAE, and IoU. Both quantitative and qualitative analysis indicate that the proposed model demonstrates exceptional capability in polyps segmentation and has great potential clinical significance. We have shared our code on anonymous website for evaluation. 
\end{abstract}    
\section{Introduction}
\label{sec:introduction}
In recent years, deep learning has revolutionized a wide range of medical imaging applications, e.g., tumor segmentation\cite{havaei2017brain,pereira2016brain}, nodule detection \cite{nasrullah2019automated}, low-dose CT denoising \cite{fan2019quadratic,wang2023ctformer,9996422,wang2023lomae}, medical imaging reconstruction \cite{xia2023diffusion,li2020xray}, etc. Among them, medical image segmentation is one of the most important research directions. Its purpose is to segment key objects in the medical image and extract effective features from the segmented areas to make the anatomical or pathological structure changes in the image clearer, thus greatly improving diagnostic efficiency and accuracy \cite{11}. In the past decades, deep learning-based technologies have been successful in a variety of medical imaging segmentation problems, including liver and liver tumor segmentation, brain tumor segmentation, video disc segmentation, heart image segmentation, and so on \cite{27,28,29,32,33}. In recent years, researchers began to pay more and more attention to the direction of polyps segmentation, which is significant because polyps, abnormal accretions on the intestinal mucosa, have a certain chance of gradually evolving into fatal cancer. Specifically, Polyps segmentation refers to identifying and labeling polyps areas from endoscopic images, which can help doctors make accurate diagnoses and treatment decisions. Accurate segmentation is very important because early identification of polyps can reduce the overall mortality caused by cancer \cite{31}. 

In recent years, several groundbreaking deep learning models have been proposed in this field and made guiding contributions to the follow-up research. Ronneberge et al. \cite{10} first proposed UNet, a U-shaped neural network formed by a convolutional network-based encoder and a symmetric decoder with shortcut connections. Zhou et al. \cite{12} further put forward UNet++, which improves UNet with more shortcut connections. They nested different sub-networks in the encoder and decoder of UNet and connected them through a series of dense hop connections, thus reducing the semantic distance and facilitating the feature inference. The important innovations of UNet and UNet++ models aim to extract richer multi-level features of images with convolution and pooling operations and transmit contextual information by cross-level feature fusion. However, their design lacks the emphasis or attention on the interested areas like the polyps, thus suffering from compromised performances.

\begin{figure*}[htbp]  
\centering
\includegraphics[width=1\linewidth]{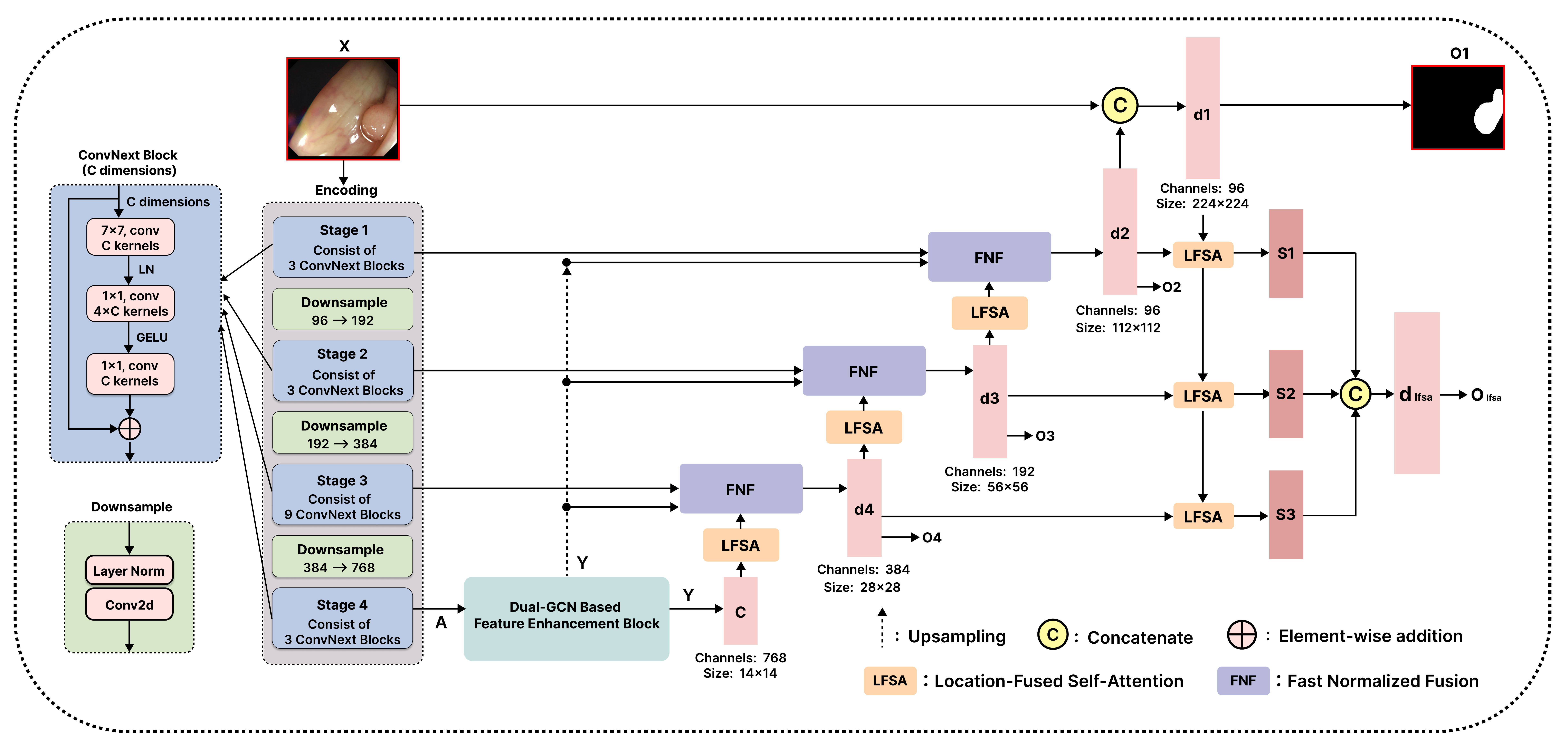} 
\caption{Overall frameworks of DSFNet.} \label{structure}
\end{figure*}

\textbf{Attention mechanism}: In recent years, transformer models have emerged as foundational architectures, particularly gaining prominence with the advent of large language models such as ChatGPT\footnote{OpenAI. (2023), ChatGPT [Large language model]. https://chat.openai.com}, LLAMA\cite{touvron2023llama}
, Falcon \cite{zxhang2023falcon}, etc. Within the transformer model, the attention mechanism is the pivotal component that can encode extensive perceptual fields through a dynamic parameter space. This transformative mechanism has found widespread applications in the area of medical imaging segmentation.
For example, Jin et al. \cite{35} combined the attention mechanism with UNet for the first time and proposed RA-UNet for CT image segmentation of liver tumors, and achieved promising results. Oktay et al. \cite{13} put forward Attention U-Net, a model that introduces an attention gate into U-Net. The attention gate can automatically learn and pay attention to the shape and size of the detection target, and pay less attention to irrelevant areas, thus improving the sensitivity and accuracy of the model. Rajamani et al. \cite{14} proposed another convolutional neural network based on attention, named AA-U-Net. They added the attention enhancement convolution module to the bottleneck section of UNet, thus enabling a more accurate spatial aggregation of contextual information. These models have been demonstrated to be effective in medical image segmentation. However, these models only pay attention to the spatial information of input images and ignore the structural information that encodes the dependence between different anatomical regions. 

\textbf{Graph convolution network (GCN)}: GCN is a special type of neural network that can encode the interrelations among each characteristic node\cite{19}. In a GCN, because each node will be updated by combining the information of itself and its neighboring nodes, the features learned by the model will include both semantic and structural features \cite{20}. Since its invention, GCN has made many breakthroughs in the field of image segmentation. Lu et al. \cite{6} put forward the convolution neural network of image semantic segmentation combining GCN and fully convolutional networks (FCN) for the first time. This model transforms the input images into graphs, thus transforming the problem of semantic segmentation into the problem of graph node classification. This method is demonstrated to surpass the performance of the traditional FCN-based methods. Hong et al. \cite{16} proposed a model combining convolution neural network (CNN) and GCN to capture the spectral characteristics of images and improved hyperspectral image classification. The structural information of the input image is also of great significance for medical image segmentation as the it can make the boundary between tissues and organs clearer. Therefore, it can relieve the edge blur problem and improve the accuracy and robustness of semantic segmentation. For example, Meng et al. \cite{37} successfully combined CNN with GCN and improved the marginal regression problem of biomedical images. Meng et al. \cite{23} later proposed a CNN based on GCN, refered to as BI-GCN, which can enhance both global semantic information and local spatial boundary features. The network has been demonstrated to have good segmentation performance on polyps and color fundus datasets under colonoscopy. Huang et al. \cite{38} proposed a new medical image segmentation neural network architecture named AGNet, also based on GCN. It successfully combines convolution features and Gabor features from shallow to deep to achieve better feature enhancement. However, even though these models have successfully extracted semantic and structural features from images using GCN, they have not paid enough attention to spatial features, which hinders the model’s ability to understand the relative relationships between objects in the image.

In this paper, we propose a novel Dual-GCN and Location-fused Self-attention model with Weighted Fast Normalized Fusion for Polyps Segmentation for polyps image segmentation. Our model leverages the benefits of Dual-GCN and a new self-attention mechanism to effectively capture both spatial and structural features of the input image. Specifically, we design a feature enhancement block module based on Dual-GCN, which can enhance the local spatial and structural information with fine granularity. To the best of our knowledge, we are the first to apply the shortest path graph attention mechanism to medical image segmentation. Moreover, we design a location-fused self-attention module to enhance the global information integration of the feature map. Finally, we introduce a fast normalized fusion method with trainable weights to efficiently integrate the outputs of the aforementioned multiple stages.

In summary, our contributions are fourfold:

\begin{itemize}
    \item We introduce a new U-shape alike network for polyps segmentation. Compared with existing models, our model design can facilitate richer and more diverse contextual information fusion with multi-scale feature aggregation in decoder. This capability allows for a more accurate delineation of the authentic boundary of polyp structures.

    \item To further advance feature extraction, we propose a feature enhancement block module based on Dual-GCN, strategically integrated into the bottleneck of the U-shaped network. This marks the first application of the shortest path graph attention mechanism in medical image segmentation, showcasing the pioneering nature of our approach.

    \item To promote the model's decoding capacity, we propose a location-fused self-attention module. This addition serves to elevate the integration ability of global information within the decoder, thereby enhancing overall segmentation performance.
    

    \item The fast normalized fusion method with trainable weights is introduced to efficiently fuse the corresponding feature maps from the encoder, bottleneck, and decoder, thus promoting information consistency and reducing the semantic gap between different stages.
\end{itemize}

\section{Method}


In this part, we will introduce the proposed model in detail. Specifically, the model can be roughly divided into encoder and decoder. In the encoder, we use ConvNext block as the backbone feature extraction network of the encoder. In addition, we introduce the Dual-GCN based feature enhancement block (DBFEB) at the end of the encoder to enhance the features of the deep feature map in both spatial information and structural information. The DBFEB module can construct a graph for the channels in each layer, so as to capture the structural relationship containing rich semantic information. This can obviously promote the task of medical image segmentation. In the decoder, we use the stand-alone self-attention module as the attention mechanism to make the model pay more attention to subtle features. Since polyps structures are generally small in medical images, adding a self-attention mechanism can potentially improve the accuracy of small target detection and improve the edge blur problem. 

\subsection{ConvNext Blocks}
While transformer models have showcased remarkable performance surpassing convolutional models, there has been ongoing scrutiny among researchers regarding the source of this superiority. Some argue that the substantial kernel size associated with expansive perceptual fields is the deterministic factor that contributes to the transformer's extraordinary capability \cite{liu2022convnet,ding2022scaling}. Given this, we are naturally curious about the potential benefits of introducing such large kernel models to address segmentation challenges and enhance model efficacy.

In our pursuit of leveraging the advantages of large kernel models, we incorporate the ConvNext module \cite{liu2022convnet} into our model since it's the first-of-its-kind large kernel model in this field. The detailed structure is depicted in the Figure \ref{structure}. Particularly, the number of channels in the input feature maps of the four stages are 96, 192, 384, and 768, respectively, and the number of channels between stages is adjusted through the downsampling layer. The four stages are composed of 3, 3, 9, and 3 ConvNext blocks, respectively. Assuming the input is $x\in \mathbb{R}^{C\times H\times W}$, it passes through a convolution layer containing the number of  $7\times7$ filters, and then a normalization layer named Layer Normalization (LN). Then, a convolution layer containing the number of $4\times C$ ($1\times1$) filters expand the number of channels of the feature map to $4\times C$ obtain deeper information. Subsequently, after passing through the Gaussian Error Linear Units layer (Gelu), the feature graph will restore the number of channels to $C$ through a convolution layer with $C$ filters. Finally, the initial input $x$ and output features are residually connected as the final output of a block. Because the essence of residual connection is an addition operation, a block will not change the number of channels of input features.

Distinct from traditional CNN which usually employs $3\times3$ kernel, ConvNext, resembling typical Swin transformer design \cite{liu2021Swin}, utilizes a larger convolution kernel of $7\times7$ to expand the receptive field of the model, thus achieving more favorable performance at various computer vision tasks than traditional CNN. At the same time, ConvNext also replaces Batch Normalization and ReLU activation functions with Layer Normalization and GELU, further improving the network’s stability, convergence speed, and generalization ability.

\subsection{Dual-GCN Based Feature Enhancement Block (DBFEB)}
\subsubsection{Spatial and Structural Graphs}

In recent years, GCN have gained widespread popularity across various domains, owing to their remarkable ability to model intricate relationships among neighboring nodes. Recognizing the dispersed and complex nature of polyp structures across different kernel areas, we find GCN to be an ideal candidate for addressing the polyps segmentation challenge. The utilization of graph relations enables capturing spatial connections between diverse anatomical regions. To leverage this potential, we introduce a feature enhancement block based on Dual-GCN, tailored for the simultaneous extraction of both spatial and structural features from endoscopic images. Specifically, the spatial graph will encode the relation tiny structures at pixel level, while the shortest path structural graph will be able to grasp long contextual interactions across anatomical regions. 


\begin{figure}[htb]
\begin{center}
\includegraphics[width=.85\linewidth]{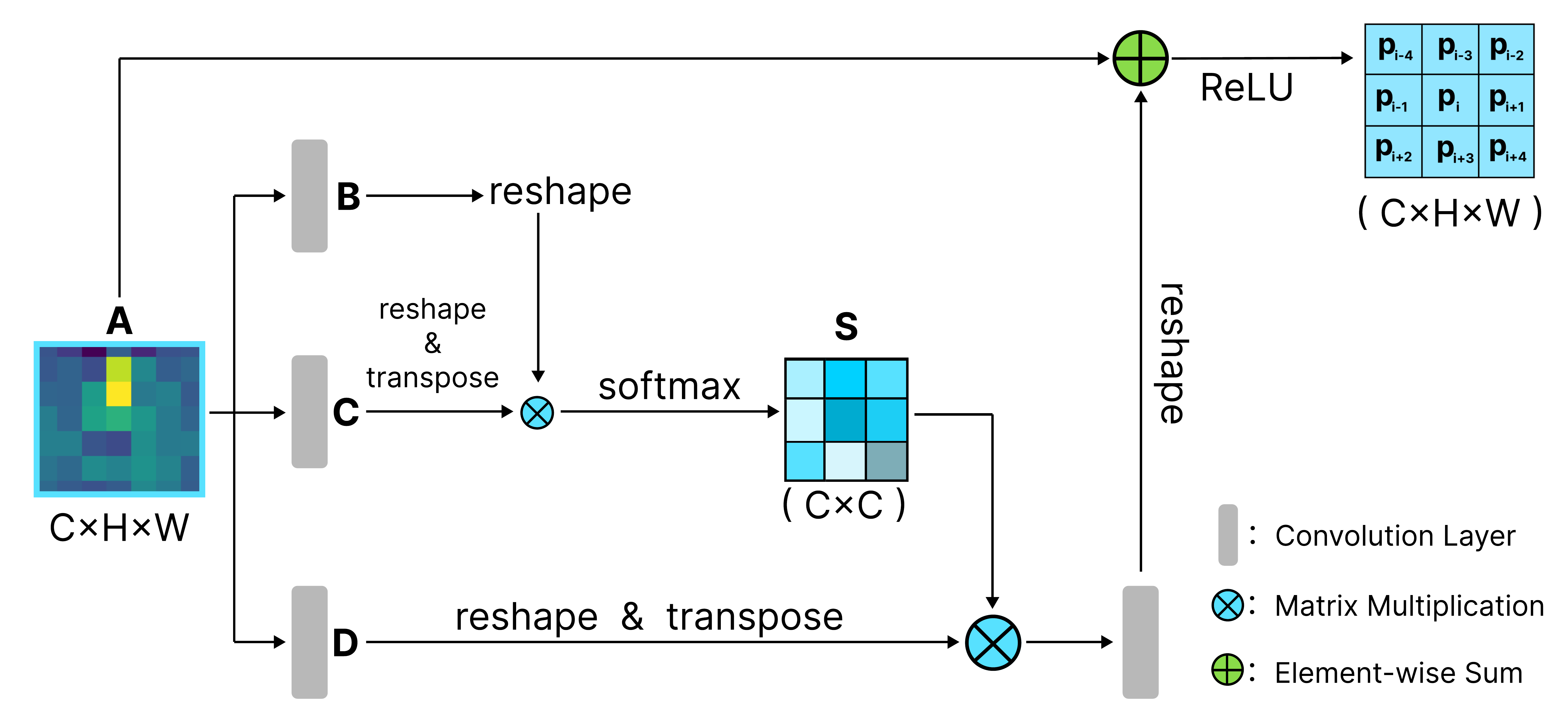}
\caption{The process of generating spatial graph}
\label{gcn02}
\end{center}
\end{figure} 

\begin{figure*}[htb]
\begin{center}
\includegraphics[width=.8\linewidth]{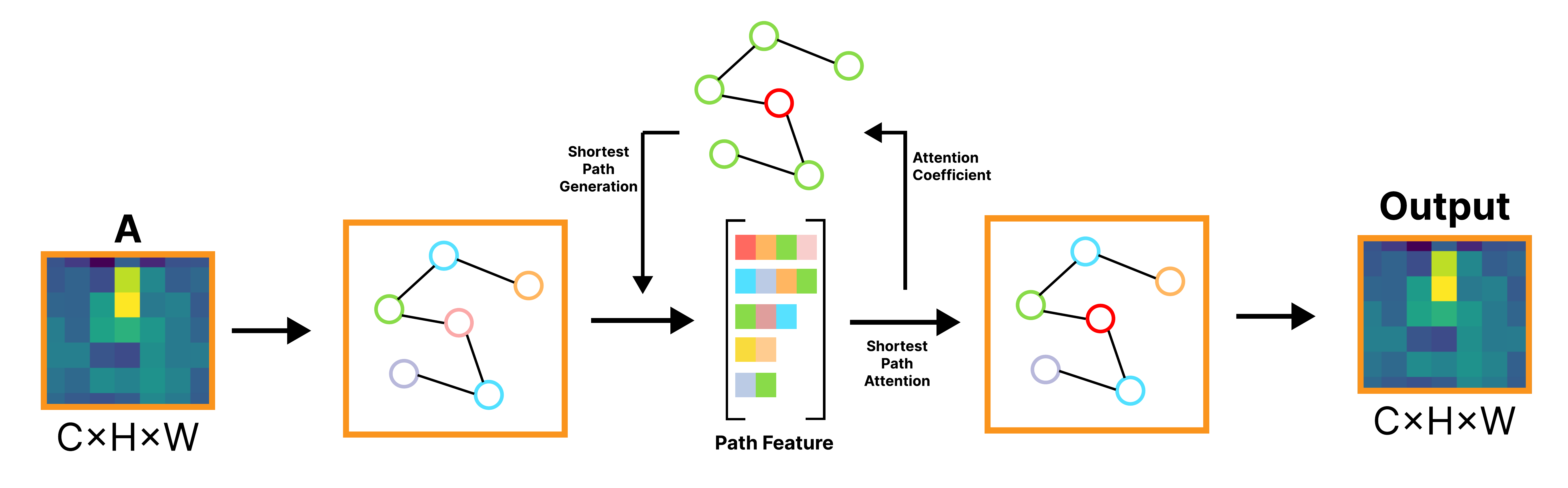}
\caption{The process of shortest path based graph attention mechanism}
\label{gcn03}
\end{center}
\end{figure*}

\textbf{Spatial graph} measures the similarity between the central pixel and its neighbors, which can provide fine spatial relationship information \cite{5}. As shown in the above Figure \ref{gcn02}, feature graphs $A\in \mathbb{R}^{ C\times H \times W}$ are first input into two convolution layers separately and generate new feature graphs $B$ and $C$, that is $\left\{B, C\right\} \in \mathbb{R}^{C\times H \times W}$. Then, we reshape $B$ and $C$ as $\mathbb{R}^{C\times N}$ where $N = H\times W$ denotes the number of pixels. Next, we feed the result of transpositional multiplication of $B$ and $C$ into softmax layer to get spatial attention map $S_p(A) \in \mathbb{R}^{N\times N}$:
\begin{equation}
S_p(A)_{ij} = \frac{exp(B_i\cdot C_j)}{\sum_{i=1}^N exp(B_i\cdot C_j)}. 
\end{equation}

Among them, $S_p(A)_{ij}$ shows the influence of the $i^{th}$ pixel on the $j^{th}$ pixel. Meanwhile, the input feature map $A$ is fed into the third convolution layer to generate a new feature map $D\in \mathbb{R}^{C\times H \times W}$. Multiply $D$ with $S_p(A)_{ij}$ after reshaping and transpose operations. Finally, the multiplied result is reshaped and added to the input feature graph $A$, and the final spatial graph is obtained after passing through a ReLU layer.

\textbf{Structural graph} can pay more attention to the structural information, which includes inter-relation between different semantics. As shown in Figure \ref{gcn03}, We use a graph attention mechanism based on the shortest path to extract the global structure information of the input feature graph\cite{yang2021spagan}. The intuition is to enable each node to aggregate the information from other distant nodes that are connected by the shortest paths, and also to explore the potential topology information of the graph. The figure illustrates the process. First, we convert the input feature graph into a graph structure. Then, for each central node, we compute a set of shortest paths to its higher-order neighboring nodes using Dijkstra's algorithm\cite{dijkstra2022note} and extract their features as path features. Next, we apply the shortest path attention mechanism to calculate the attention coefficients of each higher-order neighbor node with respect to the central node.
The edge weight from node $i$ to node $j$ can be derived by the following function\cite{yang2021spagan}:
\begin{equation}
    \mathcal{W}_{i j}=\frac{1}{K} \sum_{k=1}^{K}{ } \alpha_{i j}^{(k)},
\end{equation}
where, $\mathcal{W}_{i j}$ denotes the weight of the edge from node $i$ to node $j$, $K$ means the number of attention heads, $\alpha_{i j}^{(k)}$ means the attention weight from node $j$ to node $i$ in the $k$-th attention head. This way, we can effectively capture more structural information of the graph with the shortest path based global attention mechanism.

Different from the traditional CNN, the application of GCN can effectively transform the task of image segmentation into the problem of graph node classification. Moreover, feature fusion based on spatial and structural graphs can provide more attention to local spatial and local structural information of feature graphs at the same time \cite{6}. The specific formula for the fusion of the two feature maps is as follows $:$

\begin{equation}
    Y = f(S_p(A)\otimes S_t(A)),
\end{equation}
where, $A$ and $Y$ denote the DBFEB module input and output, respectively. $S_p(\cdot)$ and $S_t(\cdot)$ means projection of spatial graph and structural graph. $f(\cdot)$ is used to denote a layer consisting of a convolutional layer and a batch normalization layer. $\otimes$ denotes concatenation.


\subsection{Location Fused Self-Attention (LFSA)} \label{lfsa}
As an important mechanism of token learning, self-attention has been widely used in the transformer models \cite{8}. Inspired by different attention mechanisms, here, we proposed a novel self-attention block, named location fused self-attention (LFSA) to the machine learning community. Particularly, LFSA is an independent plugin-and-play self-attention mechanism that enlarges the parameter space of the attention keys, thus potentially enhancing the model's performance. Specifically, the process of building a new feature map using the LFSA module can be expressed by the following formula:
\begin{equation}
\begin{split}
y_{ij} = \sum_{a,b\in N_k(i,j)}softmax_{ab}(q^T_{ij}\cdot (\frac{\omega_{1}}{\varepsilon+\omega_{1}+\omega_{2}} \cdot k_{ab}\\
+ \frac{\omega_{2}}{\varepsilon+\omega_{1}+\omega_{2}} \cdot r_{a-i,b-j}))\cdot v_{ab},
\end{split}
\end{equation}
Here, $y$ represents a new feature map processed by LFSA model. $q,k,v$ represent queries, keys, and values, respectively. $q_{ij}=W_Q\cdot x_{ij}$, $k_{ab}=W_k\cdot x_{ab}$ and $v_{ab}=W_V\cdot x_{ab}$. $X$ represents an input image. $W_Q, W_K, W_V \in \mathbb{R}^{d_{out}\times d_{in}}$ are all learned transform. Different from the traditional self-attention, LFSA added two variables, row offset $a-i$ and column offset $b-j$, to express the relative distance of $ij$ to each position $ab \in N_k(i,j)$. $\omega_{1}$ and $\omega_{2}$ are both trainable parameters. $\varepsilon$ is a very small constant to avoid zero denominator. 


The whole process is expressed as follows:
\begin{equation}
d_{lfsa} = S_1\otimes S_2 \otimes S_3,
\end{equation}
Here, $d_{lfsa}$ is the final output and $S_i$ is calculated as:
\begin{equation}
S_i = \mathrm{LFSA}(d_i,d_{i+1}),
\end{equation}
\begin{equation}
\mathrm{LFSA}(d_i,d_j) = y(q_i, k_i, v_j).
\end{equation}




\subsection{Weighted Fast Normalized Fusion}

The proposed DBFEB demonstrates proficiency in extracting rich spatial and structural features. However, a critical challenge arises in seamlessly fusing these features with corresponding maps during the feature inference process. In response to this challenge, Tan et al. \cite{9} introduced the fast normalized fusion (FNF) method, aiming for effective feature fusion. Nevertheless, the original method lacks consideration for the distinct contributions from different feature maps.

To overcome this limitation, we present a weighted fast normalized fusion (FNF) method, which takes into account the varying contributions from different feature maps. This enhancement results in a more efficient and faster fusion process, allowing for the assignment of trainable weights to each feature map. The following equation outlines this refined procedure:

\begin{equation}
\begin{split}
O = & \sum_{k=1}^3 \frac{\omega_k}{\epsilon + \sum^6_{j=1}\omega_j}\cdot I_k \\
& + \frac{\omega_4}{\epsilon + \sum^6_{j=1}\omega_j}\cdot \frac{I_1+I_2}{2} \\
& + \frac{\omega_5}{\epsilon + \sum^6_{j=1}\omega_j}\cdot \frac{I_1+I_3}{2} \\
& + \frac{\omega_6}{\epsilon + \sum^6_{j=1}\omega_j}\cdot \frac{I_2+I_3}{2},
\end{split}
\end{equation}
In this formula, $\epsilon = 0.0001$ is a very small constant to avoid zero denominator. $I_k$ represents the $k^{th}$ fused feature map, and $I_k\in \mathbb{R}^{C\times H \times W}$.


\subsection{Loss Function}
For model optimization, we utilize the weighted sum of BCE and Dice Losses, which can stabilize the training process and alleviate the imbalance between positive and negative samples. The specific formula is computed as follows:
\begin{equation}
\mathrm{Loss}(pred,Y)=\sum_{i=1}^4 \mathcal{L}(O_i, Y) + \mathcal{L}(O_{lfsa},Y),
\end{equation}
Wherein, $d_i$ represents the feature map generated by the $i_{th}$ layer of the model decoder. $Y$ is the label. $\mathcal{L}$ indicates the weighted sum of $BCE$ and $Dice$ losses, and the weight coefficient is set to $0.5$ by default. $O_{lfsa}$ denotes the fused results from the weighed fast normalized fusion process. 


\begin{figure*}[htb]
\begin{center}
\includegraphics[width=.9\linewidth]{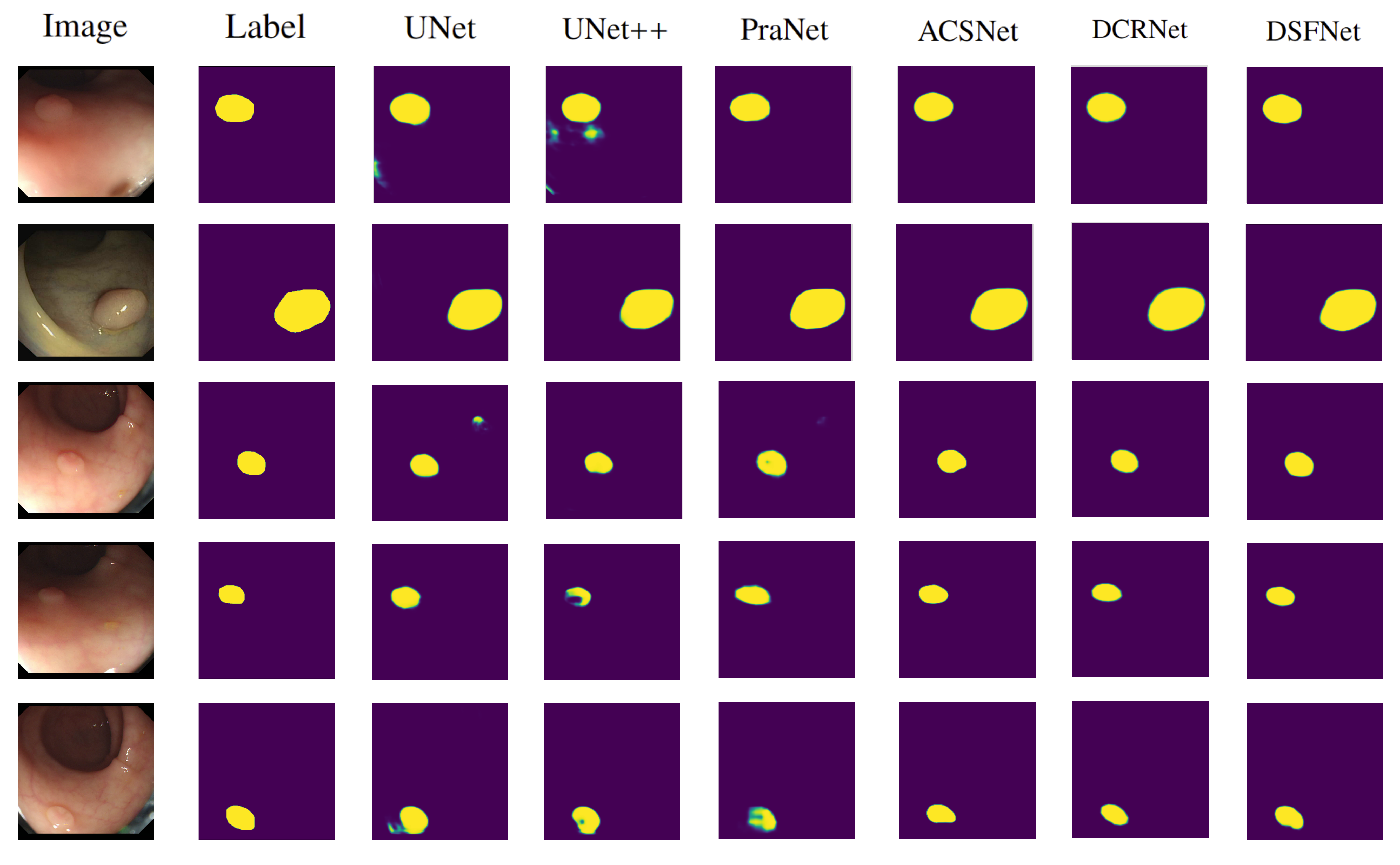}
\caption{Qualitative results of six different models on Endoscene dataset}
\label{Comparison}
\end{center}
\end{figure*} 

\begin{table*}[!htb]
\caption{Result Comparison on the Endoscene and Kvasir-SEG Dataset. The best results are bold-faced.}
\begin{center}
\scalebox{1}{
\begin{tabular}{ccccccc} 
\hline 
Dataset & Model & Dice & IoU & MAE & Boundary\_F & S\_measure\\
\hline
\multirow{6}{*}{Endoscene} & U-Net\cite{10}  & $73.78\%$ & $66.54\%$ & $4.40\%$ & $68.78\%$ & $83.54\%$\\
& U-Net++\cite{12}  & $72.88\%$ & $64.58\%$ & $4.50\%$ & $63.68\%$ & $82.41\%$\\
& PraNet\cite{47}  & $81.73\%$ & $74.38\%$ & $3.50\%$ & $75.79\%$ & $88.00\%$\\
& ACSNet\cite{48}  & $85.15\%$ & $78.67\%$ & $3.00\%$ & $81.58\%$ & $90.54\%$\\
& DCRNet\cite{49}  & $85.41\%$ & $78.86\%$ & $3.00\%$ & $83.20\%$ & $90.79\%$\\
& Ours  & $\textbf{87.83\%}$ & $\textbf{81.65\%}$ & $\textbf{2.59\%}$ & $\textbf{83.29\%}$ & $\textbf{92.05\%}$\\
\hline

\multirow{6}{*}{Kvasir-SEG} & U-Net\cite{10}  & $85.97\%$ & $78.70\%$ & $4.20\%$ & $73.13\%$ & $88.36\%$\\
& U-Net++\cite{12}  & $84.16\%$ & $76.02\%$ & $5.20\%$ & $70.33\%$ & $87.17\%$\\
& PraNet\cite{47}  & $89.20\%$ & $83.61\%$ & $3.10\%$ & $77.97\%$ & $90.96\%$\\
& ACSNet\cite{48}  & $89.32\%$ & $83.83\%$ & $3.20\%$ & $79.04\%$ & $90.96\%$\\
& DCRNet\cite{49}  & $90.14\%$ & $84.44\%$ & $\textbf{2.90\%}$ & $\textbf{82.05\%}$ & $\textbf{91.49\%}$\\
& Ours  & $\textbf{90.37\%}$ & $\textbf{84.46\%}$ & $3.20\%$ & $79.43\%$ & $91.43\%$\\
\hline
\end{tabular}
}
\label{table01}
\end{center}
\end{table*}



\section{Experiments}
\subsection{Data Description}
All experiments are carried out on two polyps datasets: Endoscene and Kvasir-SEG. 
\begin{enumerate}
    \item Endoscene is an endoscopic image dataset generated by Vázquezet al. \cite{24}, which contains two subsets: CVC-ClinicDB DB and CVC-300. Endoscene contains a total of 912 images and corresponding pixel-level labels. The image size in the dataset is 612 images($384\times288$) and 300 images($574\times 500$), respectively.
    \item Kvasir-SEG is a dataset of gastrointestinal polyp images\cite{25}. It is manually annotated by doctors and then verified by experienced gastroenterologists. Kvasir-SEG (file size is 46.2 MB) contains 1000 polyp images from Kvasir Dataset v2 and their corresponding truth values. The resolution of images contained in Kvasir-SEG ranges from $332\times487$ to $1920\times1072$ pixels.
\end{enumerate}

\subsection{Experimental Settings}
All experiments were conducted on a single NVIDIA A40 (48GB) GPU with PyTorch framework. The input images were uniformly resized to 224×224 pixels for both training and testing. For training, we employed the Adam optimizer with an initial learning rate of $10^{-5}$ and a scheduled learning rate decay. The overall training process consisted of 300 epochs with a batch size of 2. 

\subsection{Comparative Results}

The proposed model is compared with five state-of-the-art polyps segmentation models including U-Net\cite{10}, U-Net++\cite{12}, PraNet\cite{47}, ACSNet\cite{48} and DCRNet\cite{49}. All these models are implemented according to their officially disclosed codes. Particularly, five quantitative metrics are employed to fully assess the segmentation performances, which include Dice coefficient, intersection over union (IoU), mean absolute error (MAE), F1 score on boundary (Boundary\_F), and structure measure (S\_measure).

Qualitative results presented in Fig. \ref{Comparison} illustrate the efficacy of the studied methods in capturing the polyps' texture. While U-Net and U-Net++ tend to generate additional artifacts or distorted boundaries, more advanced models like PraNet, ACSNet, and DCRNet produce more accurate masks with oversmoothed boundaries. For instance, although generating good masks, DCRNet tends to lose key details crucial for diagnosis and treatment due to excessively smooth boundaries. In contrast, the proposed model stands out by generating the most accurate polyps masks, capturing the clearest and most faithful polyps structure and boundaries compared to the counterparts.

The quantitative results are summarized in Table \ref{table01}. On the Endoscene, our model surpasses all other competitors in terms of all five indicators. It's noteworthy that the proposed model outperforms the second player with a large margin of more than 2\% on Dice and IoU scores. On Kvasir-SEG data, while the proposed model performs comparably to the competitors with respect to MAE, Boundary\_F, and S\_measure, it constantly delivers the highest Dice and IoU score. In summary, the proposed DSFNet can outperform the state-of-the-art models both qualitatively and quantitatively, showcasing its potential for clinical applications.

\begin{table*}[htb]
\caption{The results of ablation study on different modules}
\centering
\scalebox{0.9}{
\begin{tabular}{cccccccc}
\hline
Dataset & Combination & ConvNext & DBFEB & LFSA & Dice & IoU & MAE\\
\hline
\multirow{5}{*}{Endoscene} & A & \checkmark & \checkmark & \checkmark & $\textbf{87.84\%}$ & $\textbf{81.65\%}$ & $\textbf{2.59\%}$ \\
& B & \checkmark & \checkmark & \ding{53} & $86.04\%$ & $78.99\%$ & $2.86\%$ \\
& C & \checkmark & \ding{53} & \checkmark & $86.75\%$ & $80.15\%$ & $2.69\%$ \\
& D & \ding{53} & \checkmark & \checkmark & $83.78\%$ & $76.40\%$ & $3.24\%$ \\
& E & \ding{53} & \ding{53} & \ding{53} & $73.78\%$ & $66.54\%$ & $4.40\%$ \\
\hline
\end{tabular}
}
\label{table02}
\end{table*}

\begin{figure}[htb]
\begin{center}
\includegraphics[width=1\linewidth]{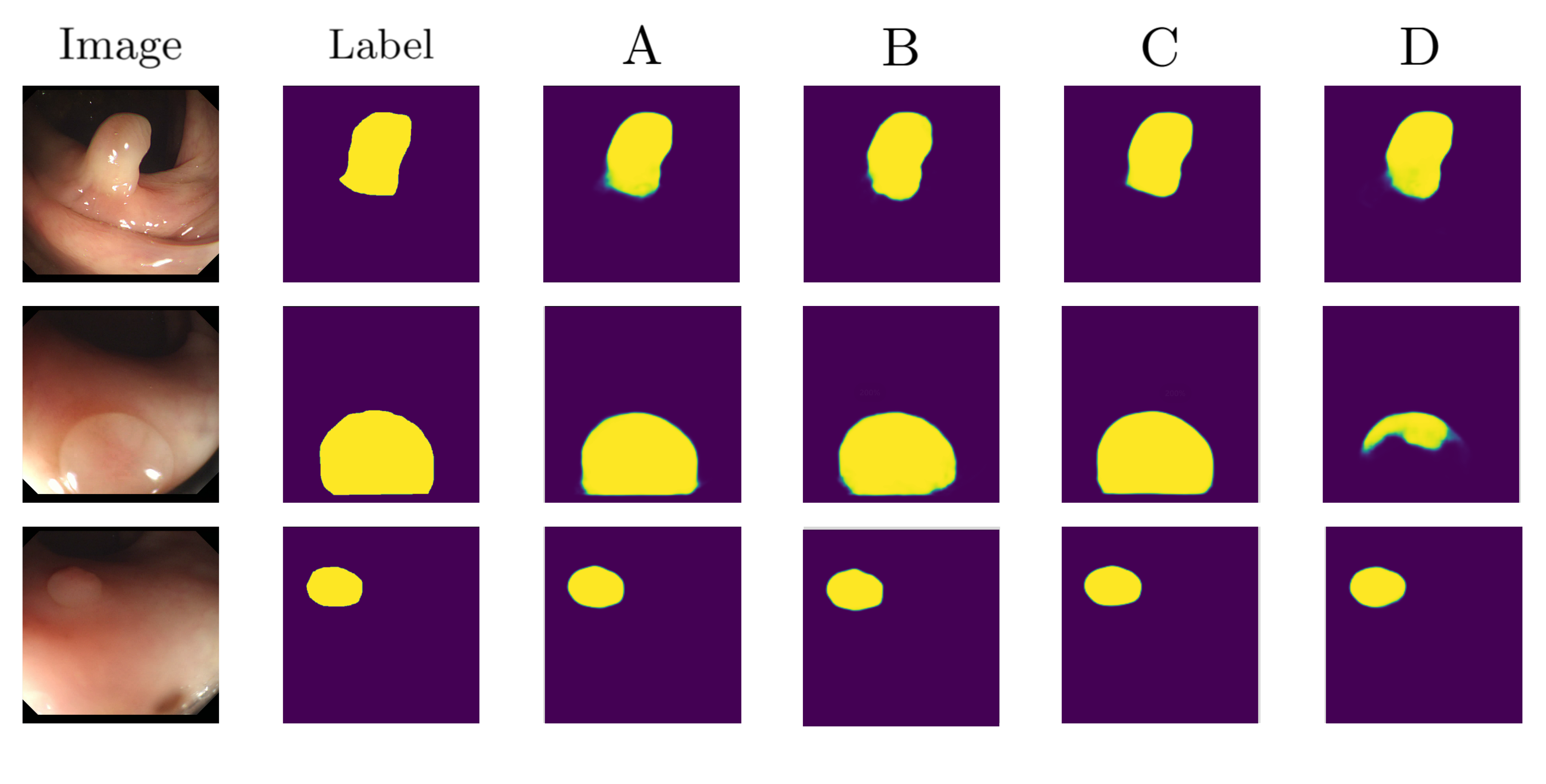}
\caption{The Figure shows qualitative results of the DSFNet with different block combinations on Endoscene Dataset}
\label{ablation01}
\end{center}
\end{figure}

\begin{table*}[htb]
\caption{Comparative results of different fusion methods and attention blocks}
\centering
\scalebox{0.9}{
\begin{tabular}{ccccccc}
\hline
Dataset & \thead{Weighted Feature \\Fusion Method} & Dice & IoU & MAE & Boundary\_F & S\_measure\\
\hline
\multirow{3}{*}{Endoscene} & \thead{UF\\} & $87.15\%$ & $80.79\%$ & $2.71\%$ & $82.59\%$ & $91.73\%$\\
& \thead{SF} & $87.32\%$ & $81.21\%$ & $2.70\%$ & $\textbf{83.59\%}$ & $91.75\%$\\
& \thead{FNF} & $\textbf{87.84\%}$ & $\textbf{81.65\%}$ & $\textbf{2.59\%}$ & $83.29\%$ & $\textbf{92.05\%}$\\
\hline

Dataset & \thead{Attention Block} & Dice & IoU & MAE & Boundary\_F & S\_measure\\
\hline

\multirow{3}{*}{Endoscene} & \thead{CCNet\cite{Huang_2019_ICCV}} & $87.51\%$ & $81.04\%$ & $2.64\%$ & $82.43\%$ & $91.74\%$\\
& \thead{CSNet\cite{10.1007/978-3-030-32239-7_80}} & $87.16\%$ & $80.49\%$ & $2.64\%$ & $81.90\%$ & $91.75\%$\\
& \thead{TransAttUnet\cite{chen2022transattunet}} & $87.53\%$ & $81.10\%$ & $2.61\%$ & $82.22\%$ & $91.65\%$\\
& \thead{Ours} & $\textbf{87.84\%}$ & $\textbf{81.65\%}$ & $\textbf{2.59\%}$ & $\textbf{83.29\%}$ & $\textbf{92.05\%}$\\
\hline
\end{tabular}
}
\label{table03}
\end{table*}

\subsection{Ablation Studies}
\subsubsection{On the impact of the key components}


Ablation studies have been conducted on all key components, including the ConvNext blocks, the DBFEB, and the LFSA modules. The results, presented in Table \ref{table02} and Figure \ref{ablation01}, reveal the insightful performance variations. Notably, the replacement of traditional CNN with ConvNext blocks yields the most significant improvement in model performance. On the other hand, the addition of the DBFEB shows the least impact on performance. However, on the Endoscene dataset, the incorporation of each module leads to noticeable enhancements in the Dice index, ranging from 1\% to 4\%. This observation underscores the substantial boosting effect achieved by introducing graph structures into the deep layers of the neural network. Leveraging these structures to simultaneously enhance the structural and spatial information has proved to be particularly impactful for the Endoscene dataset.

\subsubsection{On different fusion methods}

\begin{figure}[hptb]
\begin{center}
\includegraphics[width=\linewidth]{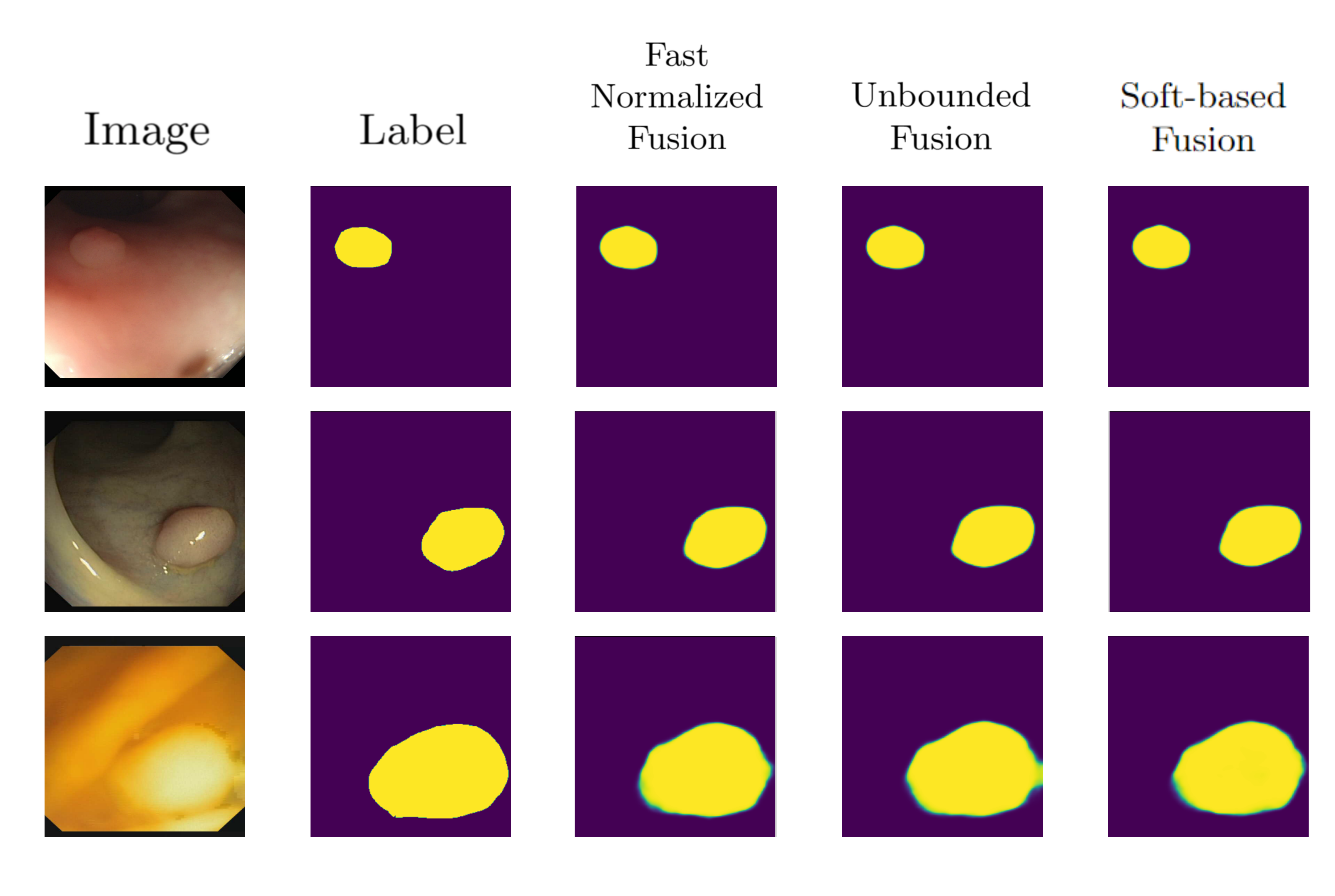}
\caption{The Figure shows qualitative results of DSFNet with different fusion methods}
\label{ablation02}
\end{center}
\end{figure}

Furthermore, as shown in Table \ref{table03} we conduct a comparative experiment with two types of weighted feature fusion methods from \cite{9}: Unbounded fusion method (UF) and Softmx-based fusion method (SF). 
UF is to add a trainable $\omega_i$ directly before each feature, and the formula is: 
\begin{equation}
O = \sum_i \omega_i \cdot I_i.
\end{equation}

SF is an improved feature fusion method which applies softmax to every trainable weight $\omega_i$, and the mathematical expression is: 
\begin{equation}
O=\sum_i \frac{e^{\omega_i}}{\sum_j e^{\omega_j}}\cdot I_i.
\end{equation}

As indicated in Table \ref{table03}, the experiment results demonstrate that the fast normalized fusion method delivers the best segmentation performance in four metrics out of five.

\subsection{On different attention blocks}
To evaluate the effectiveness of our proposed DBFEB, we conduct a comparative study with three other attention blocks that have been recently applied to medical image analysis: CCNet\cite{Huang_2019_ICCV}, CSNet\cite{10.1007/978-3-030-32239-7_80} and TranAttUnet\cite{chen2022transattunet}. We replace the DBFEB module in our network with each of these blocks and trained the models under the identical settings. As shown in Table \ref{table03}, The results indicate that our DBFEB outperforms its competitors across all metrics spanning Dice, IoU, MAE, Boundary\_F, and S\_measure.

\section{Conclusion}
In this paper, we proposed a Dual-GCN and location fused self-attention with weighted fast normalized fusion for polyps segmentation. Our model excels in capturing both spatial and structural features of polyp structures at a higher semantic level, thus enhancing the integration of global contextual information. Particularly, the model consists of three novel parts: first, the Dual-GCN feature enhancement module with the highlighted shortest path graph network was integrated into the bottleneck to enhance the feature extraction of local spatial and structural contextual information. Second, the LFSA module was incorporated to refine the global information. Finally, the weighted fast normalized fusion method with trainable weights was utilized to efficiently fuse diverse feature maps. As indicated by the experimental results on the public datasets, our proposed model demonstrates excellent segmentation performances and outperforms existing approaches in the gold standard metrics. In the future, we will explore the application of the proposed model to other medical applications. 




\bibliographystyle{unsrt} 
\bibliography{ref}

\end{document}